\documentclass[a4paper]{jpconf}
\usepackage{graphicx}
\usepackage[export]{adjustbox}
\usepackage{float} 
\usepackage{xurl} 
\usepackage[margin=1in]{geometry}
\usepackage{hyperref}
\hypersetup{
 colorlinks=true,
 citecolor=blue,
 linkcolor=blue,
 filecolor=blue, 
 urlcolor=blue,
 }

\begin{document}

\title {A novel internship program in HEP}

\author{Santanu Banerjee$^1$, Tulika Bose$^2$, Ulrich Heintz$^3$, Sudhir Malik$^4$, }

\address{$^1$Physics Department, Tougaloo College, Tougaloo, MS 39174, U.S.A.}
\address{$^2$Physics Department, University of Wisconsin - Madison, Madison, WI 53715, U.S.A.}
\address{$^3$Department of Physics, Brown University, Providence, RI 02912, U.S.A.}
\address{$^4$Physics Department, University of Puerto Rico Mayaguez, Mayaguez, PR 00682, U.S.A.}

\ead{sudhir.malik@upr.edu}

\renewcommand{\footskip}{10pt} 

\begin{abstract}
The U.S.~CMS collaboration has designed a novel internship program for undergraduates to enhance the participation of students from under-represented populations, including those at minority serving institutions, in High Energy Physics (HEP). These students traditionally face several barriers including lack of research infrastructure and opportunities, insufficient mentoring, lack of support networks, and financial hardship, among many others, resulting in a lack of participation in STEM fields. We had recently reported about a fully virtual 10-week internship pilot program called ``U.S.~CMS - PURSUE (Program for Undergraduate Research SUmmer Experience)'' to address dismantling such barriers. The 2023 iteration of this program builds on it by imparting not only an in-person summer internship experience but extends it into the academic semester as well. Students are selected predominantly from Minority Serving Institutions with no research program in HEP and from under-represented groups. They experience a structured hands-on research experience with an initial two-week ``bootcamp" on software training modules followed by an 8-week HEP project targeting physics analysis, software, computing or instrumentation work on the CMS Experiment. A subset of interns continue the experience into the academic semester, enabling a further in-depth knowledge of the field and a motivation to persist in STEM areas. In this paper, we describe our recent experience with this upgraded internship program. The paper is dedicated to the memory of Prof. Meenakshi Narain (Brown University) who was the driving force behind this internship program and U.S CMS diversity, equity and inclusion efforts.
\end{abstract}

\section{Introduction}

Students from underrepresented populations, including those at minority serving institutions, have traditionally faced barriers that have resulted in their being under-represented in High Energy Physics (HEP). These barriers include a lack of research infrastructure and opportunities, insufficient mentoring, lack of support networks, and financial hardship, among many others. To help overcome these barriers, some members of the U.S.~CMS~\cite{USCMS} collaboration designed a fully virtual research pilot program, called ``U.S.~CMS {\bf P}rogram for {\bf U}ndergraduate {\bf R}esearch {\bf Su}mmer {\bf  E}xperience" or ``PURSUE"~\cite{pursue2022}, that was successfully executed in Summer 2022. Building upon this experience combined with feedback received, a novel upgraded version was launched in summer 2023. This 10-week in-person research internship program includes a 2-week cohort-based training experience hosted at Fermilab and remaining 8-weeks at one of the U.S. CMS institutions. It also provides a mechanism whereby a subset of interns continues the experience into the academic semester, enabling a deeper dive into the field and a motivation to persist in STEM areas. The funding for this program is provided by a DOE RENEW-HEP grant: U.S.~CMS SPRINT award led by Tougaloo College along with Brown University, University of Puerto Rico, and University of Wisconsin with support from the U.S.~CMS Operations Program. Tougaloo College is a member of the Historically Black Colleges and Universities (HBCU) group that has produced a significant number of educators and STEM professionals from Mississippi.  It is  a top college for undergraduate origin of African American STEM PhDs
in Biological and Physical Sciences over a recent 20-year period~\cite{nsf_tougaloo,nih_tougaloo} and a top 20 Women Friendly Physics department in the nation~\cite{women_tougaloo}. Tougaloo College also has a long history of partnership~\cite{brown_tougaloo} with Brown University that consists of various programs that engage the culture, academics, and histories of these two distinctive institutions. The bonding between Tougaloo and U.S. CMS via the internship program  provides an avenue to expand on the experiences at Tougaloo College to attract and support under-represented minority (URM) undergraduate students nationwide and train them in High Energy Physics. This unique perspective provides an opportunity for minority students in Physics to experience the best of diverse experiences. The internship program is in line with U.S. CMS DEI plan~\cite{USCMS-DEI} and addresses the lack of diversity in physics/high-energy physics (HEP) due to racial, ethnic or gender identity by focusing on URM students including Black, Hispanic, and indigenous communities as well as other marginalized communities. Multiple U.S. CMS  institutions take part in successful execution of this internship program making it unique and one of its kind.

\section{A novel internship at the science frontier}
The experiments at the Large Hadron Collider (LHC)~\cite{lhc_cern} are one of the big drivers of science at the frontiers of human knowledge and examine the smallest constituents of our vast and complex universe at scales governed by fundamental principles of quantum physics. The physics program of the Compact Muon Solenoid (CMS)~\cite{cms_experiment} and other LHC based experiments will continue in the next two decades and preparations for future large HEP experimental endeavors are being planned in parallel. The U.S.~CMS is a collaboration of 54 U.S. universities and institutes that work on the CMS Experiment at the LHC. It the largest national group in the CMS Experiment, comprising about 30\% of its collaborators with about 1200 physicists, graduate students, engineers, technicians, and computer scientists. In addition, there are a few hundred undergraduate students who are engaged in its research program. The U.S. CMS collaboration makes significant contributions to nearly every aspect of the CMS detector throughout all phases, including construction, installation, and data-taking. It also plays a major role in the construction and operation of the experiment's computing facilities and software that is used to analyze the unprecedented amount of data that CMS generates. These highly sophisticated computing tools allow physicists to operate the CMS detector, reconstruct the data, analyze it and, ultimately, make discoveries. U.S.~CMS is in a unique position to provide pathways to involve many young researchers ranging from undergraduate, graduate, and postdoctoral researchers in every aspect of the experiment and prepare hundreds of next-generation scientists and related workforce. This internship offers opportunities to the students in state-of-art detector design and upgrades, operations, novel techniques in data taking and analysis, scientific presentations and international partnerships and leads up to the above pathways. Future discoveries and scientific advancements coming from the LHC experiments is an enormous source of motivation and excitement for future scientists. 

\section{Collaborative Ecosystem}
\label{Sec:ecosystem}
\begin{figure}[ht]
    \centering
    \includegraphics[width=0.8\textwidth]{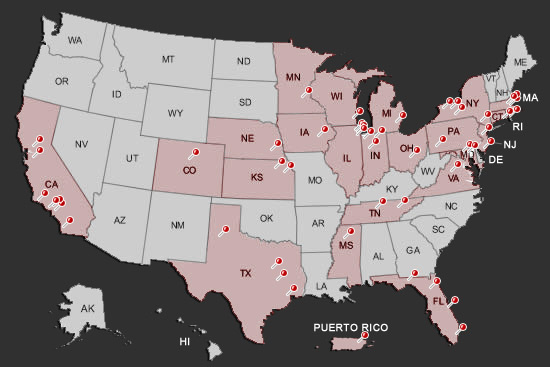}
    \vspace{0.1in}
      \caption {A map if U.S. CMS collaborating institutes mentioned in~\ref{Sec:ecosystem}.
      }
       \label{uscmsmap}
\end{figure}

The U.S.~CMS Collaboration acknowledges the strength of  diversity, equity, and inclusion (DEI) as fundamental values of our community that empowers its members to achieve their full research potential. Its highly collaborative environment is committed to increasing the representation of women and historically under-represented and marginalized groups in HEP. The previous publication~\cite{pursue2022}  outlines its DEI goals. Its unique strengths, background and pioneering leadership~\cite{uscmscontributions} in CMS and LHC and with Fermilab as the host institution in the U.S., make this internship program a rewarding experience. Fermilab LPC~\cite{lpc_fermilab} provides the computing and software infrastructure required for the internship. The first two weeks spent at Fermilab facilitates bonding of interns not only with each other but also with interns participating in other programs at Fermilab. Special accommodation in schedule is made to allow participation in all common Fermilab internship activities. The mentors for 2-week software training and 8-week projects are physicists drawn from its collaborative institutions on voluntary basis. Post Fermilab, for remaining eight weeks, interns are co-located with their mentors at various U.S. CMS institutes, a map of which is shown in Figure~\ref{uscmsmap}. This process gives a unique aspect to the internship where a diverse group of interns gets trained by a multi institutional collaboration.

\vspace{0.04in}

A subgroup of mentors and selected mentees continue to work remotely during the semester extending the summer work with a possibility to physically travel to mentor, if required, given the constraints of academic commitments of interns back at their home institute during the academic year.

\section{Internship Framework}
\label{Sec:framework}
\begin{figure}[ht]
    \centering
    \includegraphics[width=1.0\textwidth]{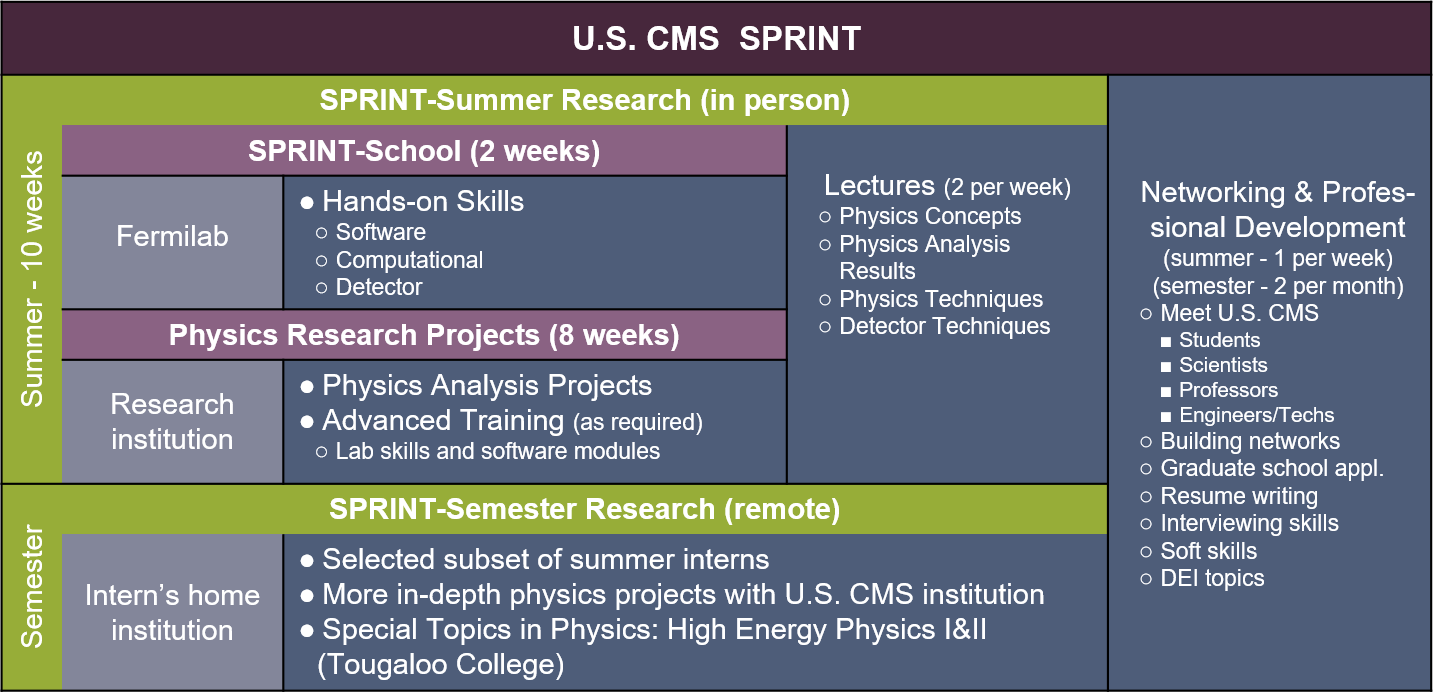}
    \vspace{-0.3in}
      \caption {A summary of the U.S. CMS SPRINT Program Structure. A strong mentor network is key to implementing the activities, and described in section~\ref{Sec:framework}.
      }
       \label{SPRINTfig}
\end{figure}

The internship is divided into two parts. The first part is an in-person 10-week summer internship and second is the semester long in-depth extension of it. The semester part cannot be taken independently as only interns selected from summer internship continue into it since it is based on summer research and related skills learnt. The first part of the internship has about 20 students of which 6 are selected to continue into the semester. The exact numbers are determined by the funding availability. The summer part is similar in structure to the pilot program and is described in detail here~\cite{pursue2022} and includes software training, physics projects, 3 lectures per week on various topics related to CMS and few evaluations. While the summer 2022 pilot internship was fully remote and attributed to emergence from the COVID-19 pandemic, the summer 2023 internship program was in-person which makes a big difference. It is not a surprise~\cite{nature} that in-person experience is critical for exchange of scientific ideas, progress and breakthroughs and in this case for peer bonding and networking as well. A graphical sketch of the U.S. CMS SPRINT program is presented in Figure~\ref{SPRINTfig}.

After advertising the program and subsequently organizing two webinars in Nov-Dec 2023, we received an excellent response from prospective interns with about 150 students applying to the program. From this set, we selected a total of 20 students who would be trained in the software tools and data analysis techniques used in HEP followed by one main project per student under the mentorship of a U.S. CMS scientist during a 10-week program in the summer of 2023. The two-week in-person training (June 5-June 17) of the 20 interns was held at Fermilab. Following the training, 8 students remained at Fermilab for the subsequent 8-week research experience, and 12 students were spread out among U.S. CMS institutions, including Purdue (3), Princeton (2), Univ. of Wisconsin (2), Univ of Alabama, Univ. of Maryland, Florida Inst. of Technology, Univ. of Nebraska-Lincoln, and Rice University. All of the interns were enrolled in a new credit bearing course (PHY 495: Special Topics in Physics: High Energy Physics) offered at Tougaloo College and received credit for this course. Laptops were acquired for the interns to provide each student with a common computing platform including relevant open-source software necessary for training and research.

\subsection{Intern selection process}The selection of applicants for the summer internship program followed an equity-minded rubric developed using holistic components including non-cognitive competencies such as the ability to work as part of a team, commitment towards long-term goals, evidence of perseverance in the face of adversity, \textit{etc}. Our eligibility requirements were designed to be minimal to ensure that a broad range of applicants felt qualified to apply. Thus, applicants were only required to be enrolled at an undergraduate institution in the U.S., submit short responses to prompts on research interests and other experiences, submit a transcript, and submit one letter of recommendation from someone familiar with their coursework or aptitude for research. Special consideration was given to students at HBCUs, HSIs, MSIs, community colleges, and other institutions that do not traditionally participate in HEP research to increase access to those students. A selection team comprising several members of the U.S. CMS DEI Committee reviewed all the applications.  About 150 students applied for the 2023 summer program and a total of 20 students were selected by this process.

Selection of interns for the semester is based on several factors and given a score based on the following: input from mentors, feedback on final presentations from peers and experts, attendance of weekly lectures. Once the selection is done, availability of mentor and mentee pair is sought to continue during the semester. Students receive credits for the semester internship from Tougaloo College.

\subsection{Selection of mentors}The selection of research mentors took place in Spring 2023. Following a call for volunteers, students and mentors were matched based on the skill set of the student, their goals, the mentor’s availability, and the nature of the project (e.g. instrumentation, software, analysis etc.). The primary research mentor was either the PI of the research group, or a group member who would be available to meet with the intern on a regular basis, giving them helpful pointers or help problem solve, such that the students made continual progress in their research project. We also asked that secondary mentors be defined by each group such that the student had somebody to seek help from in case the primary mentor was unavailable or traveling. Where possible, we tried to find multiple primary mentors at the same institute. This enabled us to place multiple students in one location allowing them to benefit from each other’s company and have a better sense of belonging. We encouraged primary mentors to integrate the student interns within their research groups and alongside their own undergraduate students such that the interns had a more cohort-like experience. 
  
\subsection{Training and professional development} Invited talks on a broad range of topics were given by U.S. CMS collaborators and Fermilab scientists. A total of 42 hours of hands-on tutorials covered various topics on software and computing skills required for HEP research.  Tours of several research facilities at Fermilab and sessions on networking, professional development and graduate school preparation were organized. Two of the sessions merit a special mention and are described below.
\subsubsection{Entering Research:} 
\begin{figure}[ht]
\centering
\includegraphics[width=35pc, frame]{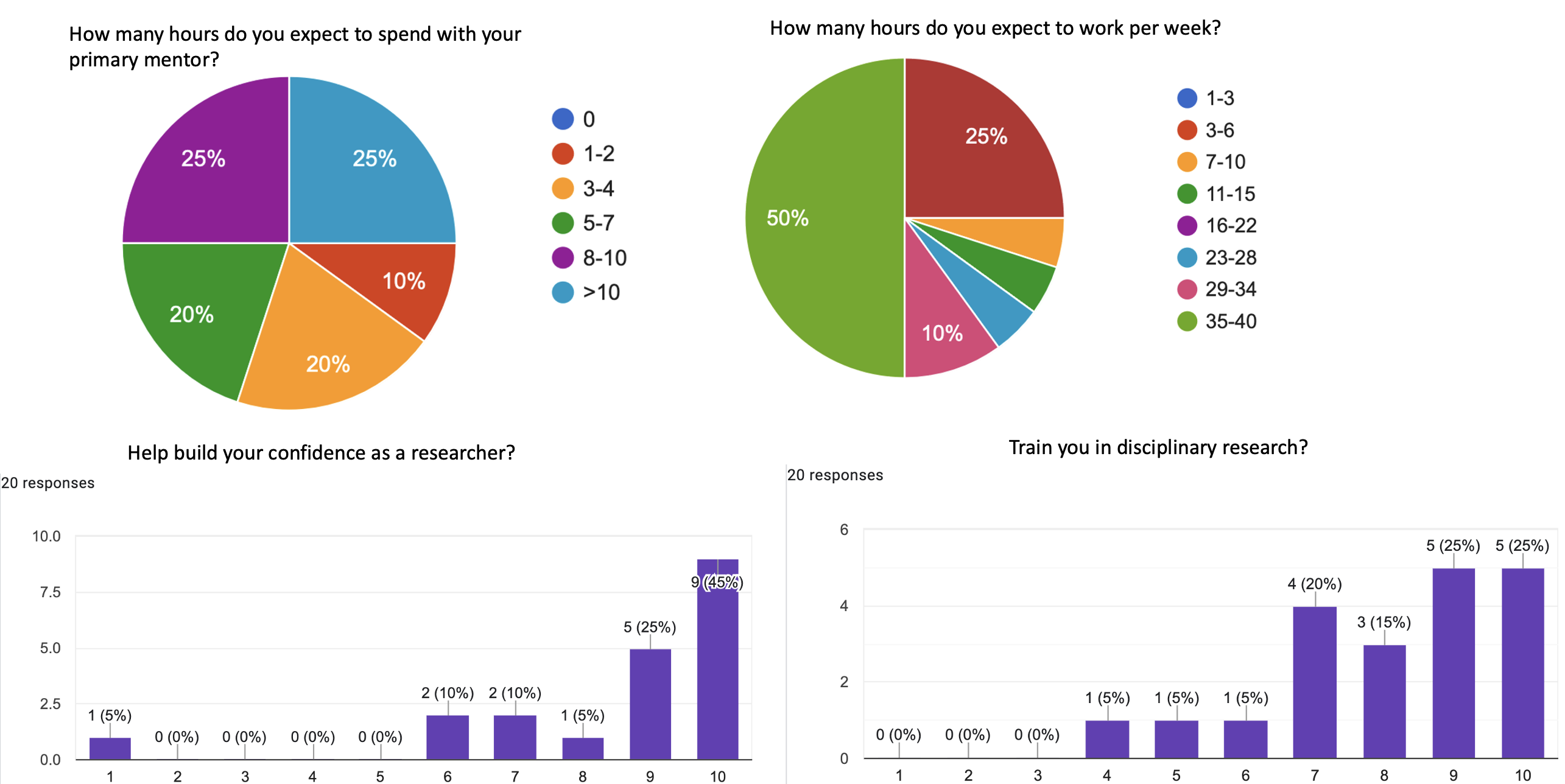}\hspace{5pc}%
\caption {Interns expectations}
\label{fig:expectations}
\end{figure}

This session was organized on the very second day of the start of the internship to set expectations as a researcher, develop identity as a researcher and develop an understanding of the research environment. Interns were asked the following questions. They were divided into groups to discuss answers and fill a feedback survey. 
\begin{itemize}
    \item What specific goals do you hope to achieve in your research experience?
    \item Why do you want to do research?
    \item What specific goals do you hope to achieve in your research experience?
    \item What are your expectations of working with a research team? 
    \item What do you think will be expected of you as an undergraduate student conducting research on a “real” research team?
    \item What contributions will you bring to your research team?
    \item What is your greatest concern, and what are you excited the most about?
    \item Explain your understanding of the scientific process as you see it today. How does a scientist approach a research question?
    \item What do you think are important abilities/skills for an individual to have to be able to conduct research?
    \item Which of those abilities/skills do you have?
    \item Which of those abilities/skills do you lack? What can you do to develop the abilities that you think you may lack?
\end{itemize}
Some expectations from the interns are shown in Figure~\ref{fig:expectations}.

\subsubsection{Entering Mentoring}On the third day of the start of the internship, a session called ``Entering Mentoring" was organised for the mentors in collaboration with the U.S. CMS DEI Committee. Material from the CIMER~\cite{cimer} project (Center for the Improvement of Mentored Experiences in Research) was used to cover a broad range of issues with emphasis on aligning expectations and maintaining effective communications with mentees.  A survey, case studies and mentor-mentee agreements formed part of this session as well.

\begin{figure}[ht]
\centering
\includegraphics[width=35pc, frame]{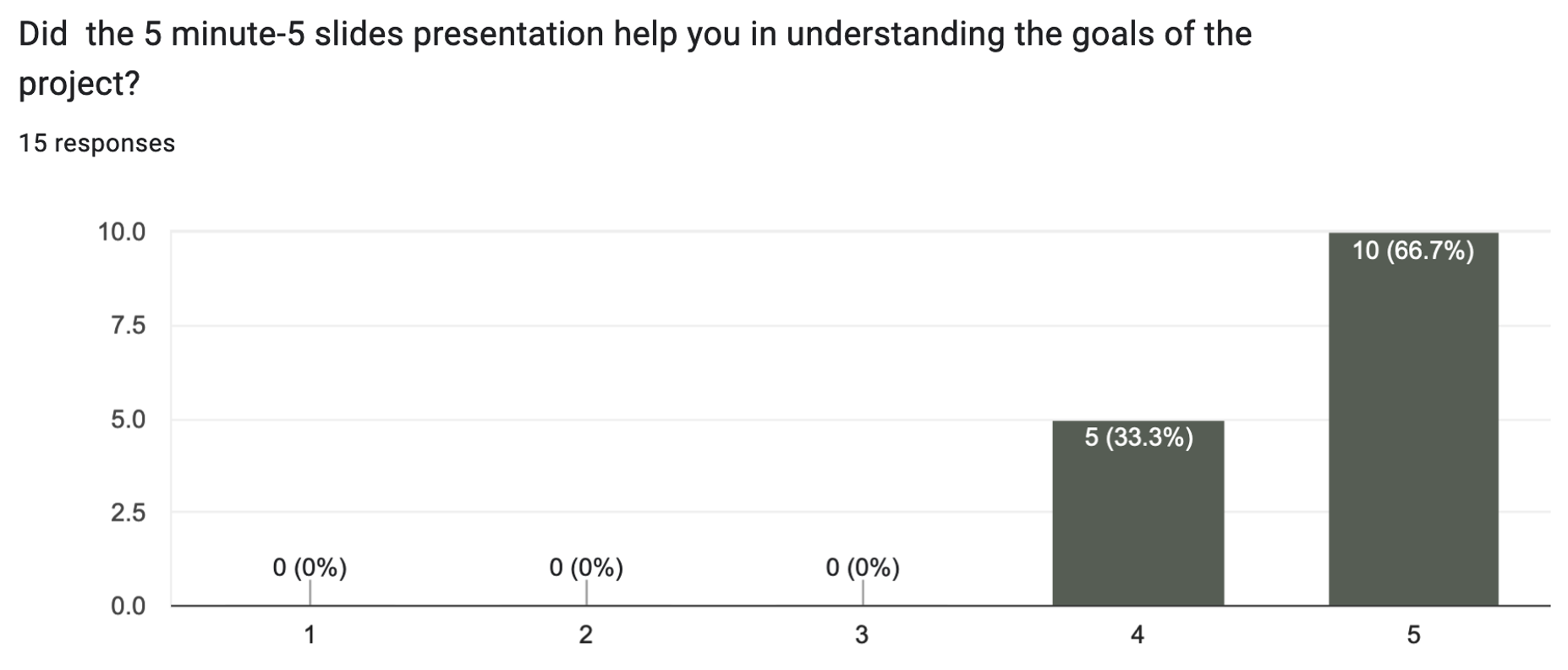}\hspace{5pc}%
\caption {5 slides in 5 minute presentations }
\label{fig:5min5slides}
\end{figure}

\begin{figure}[ht]
\centering
\includegraphics[width=35pc, frame]{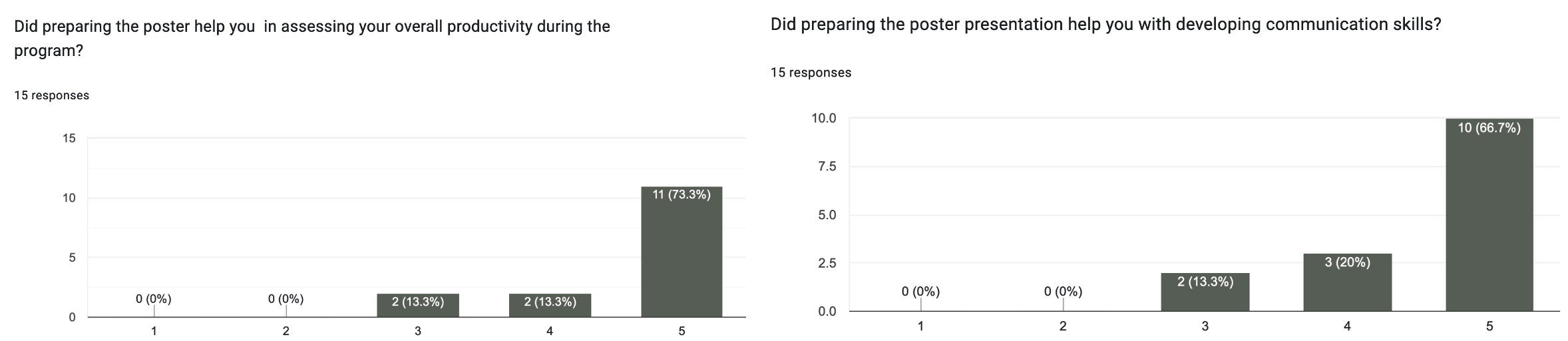}\hspace{5pc}%
\caption {Poster presentations feedback}
\label{fig:poster}
\end{figure}

During the initial weeks of the internship, interns were asked to meet with their research mentors, discuss their projects, and then present their understanding of their projects in an exercise called “Five slides in five minutes”. This presentation exercise helped the research mentors to evaluate the initial level and grasp of physics by the student, and in parallel gave the students the space to get familiar with ideas and skills they would learn during the internship and instilled in them the confidence to present their initial understanding before the experts. Figure~\ref{fig:5min5slides} shows that the usefulness of the above exercise. (Note: For Figures~\ref{fig:5min5slides}~\ref{fig:poster}~\ref{fig:onlinevsinperson}~\ref{fig:overall} and~\ref{fig:rate} the x-axis scale is 1-5 with 5 being the highest rating) 

At the end of the internship students were required to make a poster based on their physics research projects and also give short oral presentations. Figure~\ref{fig:poster} shows the responses regarding the usefulness of making a poster. Each intern’s final presentation was reviewed for feedback by their peers, mentors and reviewers. In addition, mentors provided a detailed review of the entire duration of work of their respective interns. Students were highly encouraged to make presentations at future conference opportunities. Several interns made presentations based on internship work at department symposium and local conferences such as the Gulf Coast Undergraduate Research Symposium and  the 2023 National Society of Black Physicists Conference.

\section{Assessment}
\begin{figure}[ht]
\centering
\includegraphics[width=35pc, frame]{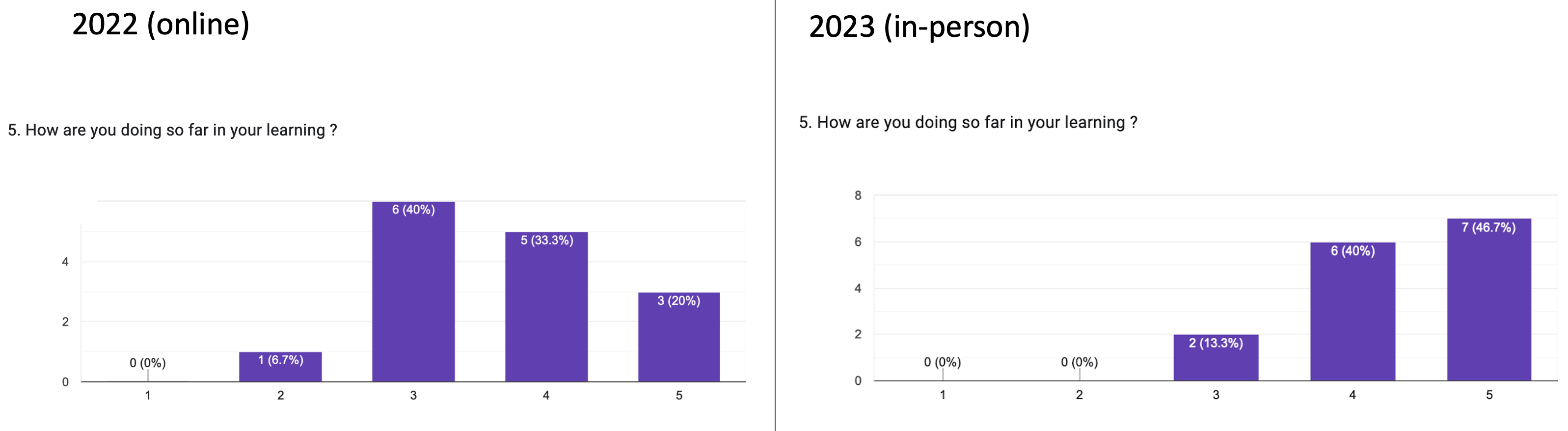}\hspace{5pc}%
\caption {A comparison of learning status three weeks into the internship: left is 2022 (online) and right is 2023 (in-person)}
\label{fig:onlinevsinperson}
\end{figure}

\begin{figure}[ht]
\centering
\includegraphics[width=35pc, frame]{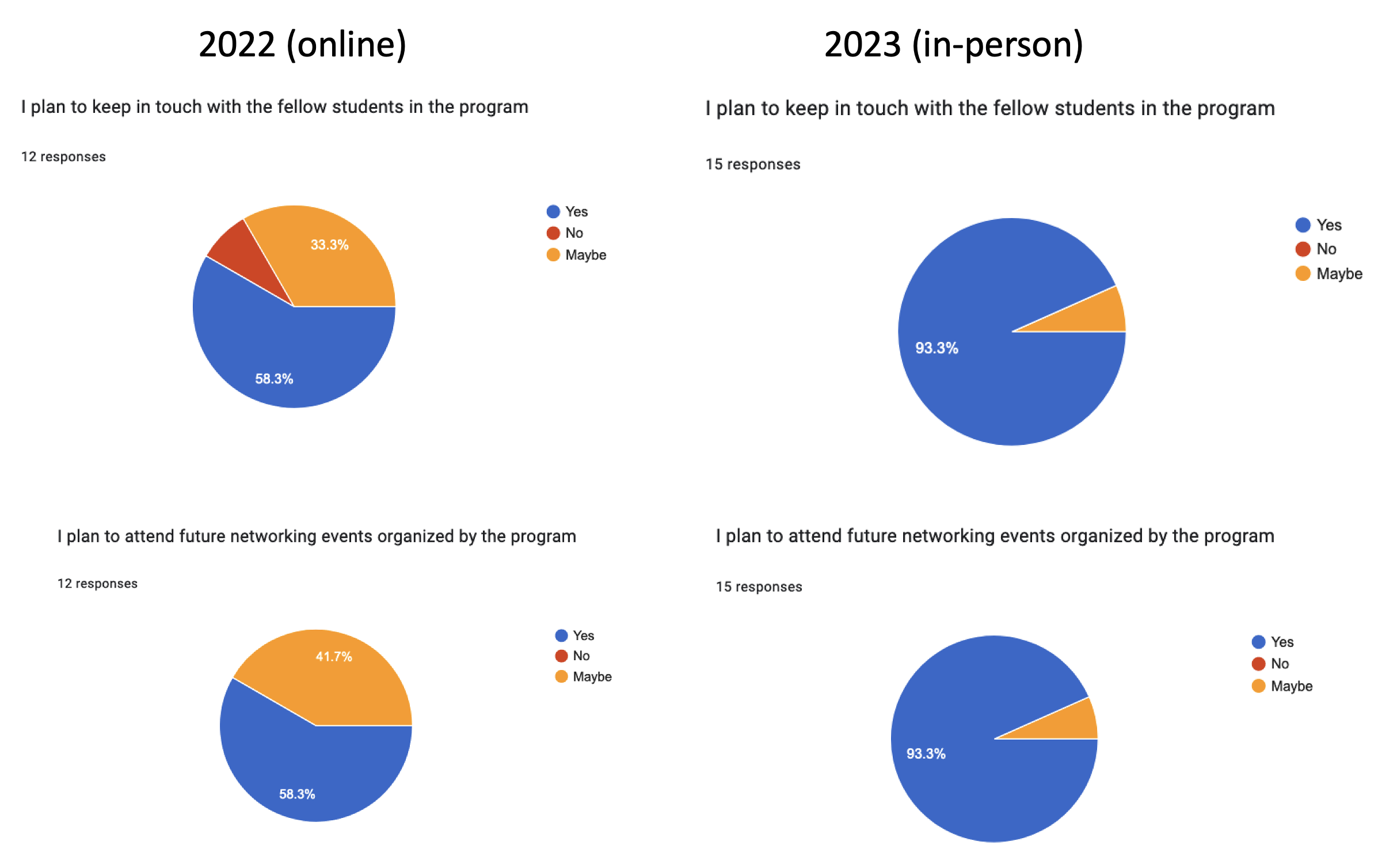}\hspace{5pc}%
\caption {A comparison of motivation to stay in touch with peers and participate in networking: left is 2022 (online) and right is 2023 (in-person)}
\label{fig:futurebonding}
\end{figure}

One very clear picture that emerged from comparative feedback is that the in-person experience in 2023 is superior to the online pilot internship in 2022. A feedback on learning status three weeks into the internship clearly demonstrates that the in-person participation is superior as shown in Figure~\ref{fig:onlinevsinperson}. A feedback at the end of internship asking about motivation to stay in touch with fellow interns and participate in future networking clearly indicates a diminished interest in case of the online internship experience as shown in Figure~\ref{fig:futurebonding}.

\begin{figure}[ht]
\centering
\includegraphics[width=35pc, frame]{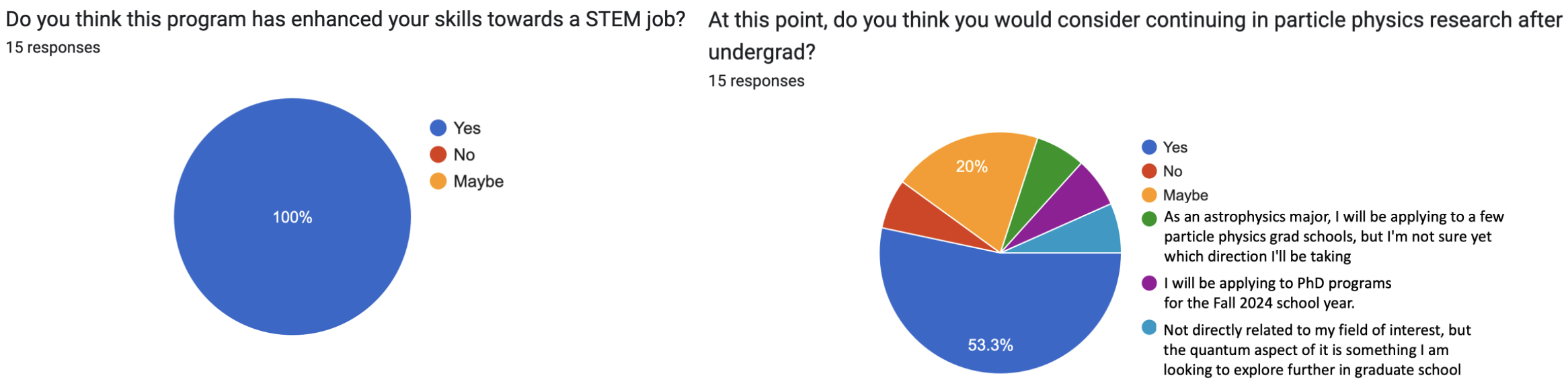}\hspace{5pc}%
\caption {All interns have skills enhanced and many want to take particle physics for grad school}
\label{fig:skillenhancedstemcareer}
\end{figure}

It is apparent that the internship experience leads to enhanced skills towards a STEM career and a majority of interns would like to pursue particle physics for research after completing their undergraduate degree as indicated in Figure~\ref{fig:skillenhancedstemcareer}. 

\begin{figure}[ht]
\centering
\includegraphics[width=35pc, frame]{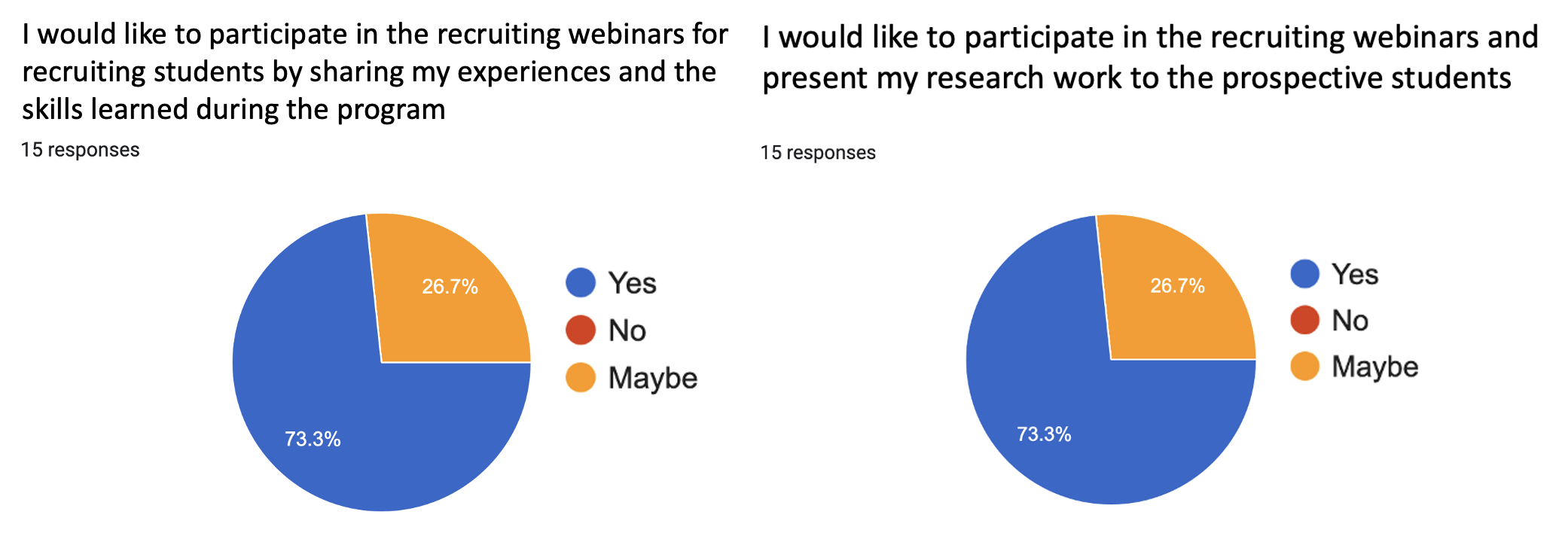}\hspace{5pc}%
\caption {Many interns would like to share experience during webinars organised for internship iterations}
\label{fig:futureparticipation}
\end{figure}

Figure~\ref{fig:futureparticipation} shows that a significant majority of interns expressed willingness to share excitement about research experience and work with prospective interns during future informational webinars about the internship. 

\begin{figure}[ht]
\centering
\includegraphics[width=35pc, frame]{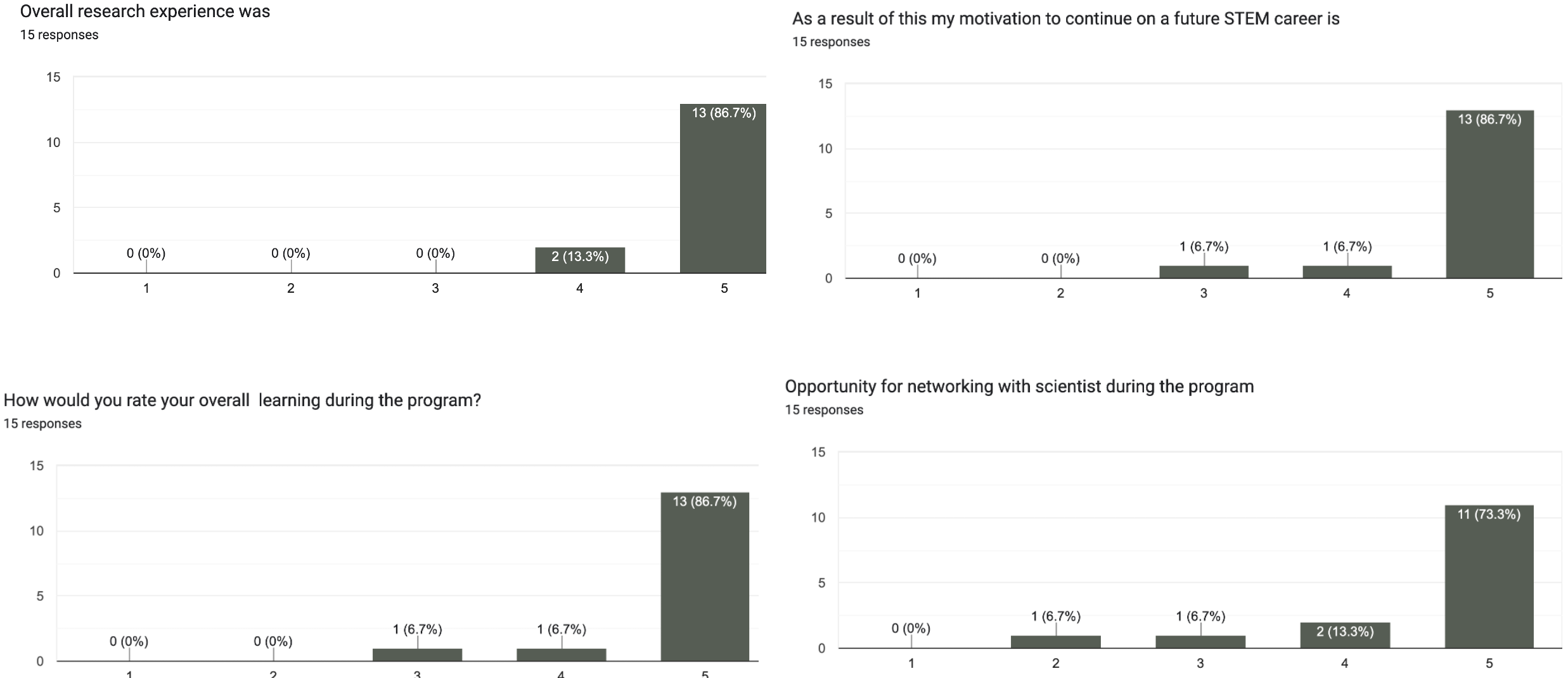}\hspace{5pc}%
\caption {}
\label{fig:overall}
\end{figure}

\begin{figure}[ht]
\centering
\includegraphics[width=35pc, frame]{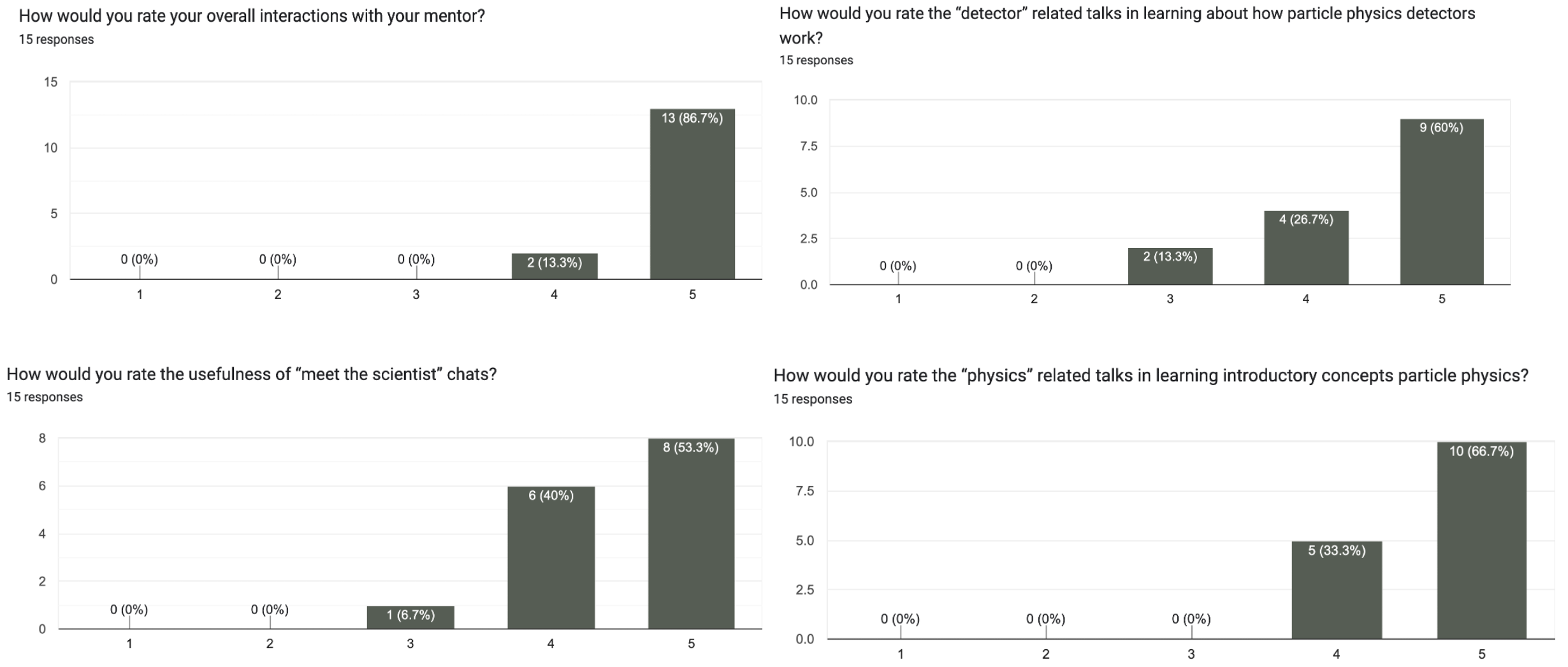}\hspace{5pc}%
\caption { }
\label{fig:rate}
\end{figure}

The overall research and learning experience feedback and motivation to stay in STEM is overwhelmingly good as shown in Figure~\ref{fig:overall}. The opportunity for networking with scientists at Fermilab, interactions with mentors and quality of talks throughout on various topics are rated very high as shown in Figures~\ref{fig:overall} and  ~\ref{fig:rate}.

A summary of main features of the internship program are enumerated below:
\begin{itemize}
    \item Developed a process to advertise amongst a diverse set of prospective students via webinars, emails to department heads, and including motivational talks by past interns from the pilot program talking about their experience. 
    \item  Implemented a selection method following a very robust rubric to select 20 interns out 150 applicants. 
    \item Involved a large collaboration (U.S. CMS) comprising 54 U.S. institutes with over 1000 people, in several aspects of the program (mentoring, training, seminars, professional development etc.) to make it a success.
    \item Trained a diverse group of students from non-research-intensive institutions in software tools, data analysis techniques and methods used in HEP research.
    \item Provided networking opportunities to the interns, amongst themselves as well as with U.S. CMS scientists (and their institutes) to broaden their cohort experience and increase their chances of pursuing a graduate school in particle physics or another STEM area. 
    \item Enabled interns to develop skills to present scientific work to a group of scientists and peers, break the barrier of hesitation and enhance confidence. 
    \item Provided dedicated mentoring sessions for the interns to bring awareness to the importance of an environment that fosters a workspace where all members can thrive and bring in their talents, irrespective of age, career status, employment situation, institutional affiliation, geographical location, nationality, gender, ethnicity, family situation, sexual orientation, or disabilities and in addition ensure a positive, respectful, and supportive climate within the U.S. CMS collaboration. 
    \item Provided dedicated resume writing sessions to help prepare portfolio for applying to PhD programs. 
    \item Provided dedicated career development sessions to create awareness of the job opportunities that take advantage of the skills developed during the internship. 
\end{itemize}    

It is also worth mentioning that the unique style of this program is demanding on the handling of logistics. Interns needed to travel and reside in Fermilab area for two weeks and then some of them travel further to their project mentor's institution for remaining eight weeks where they needed to be accommodated. It was a challenge to organise this within the constraints of budget and time. Maggie Slusarczyk (admin support at Tougaloo University) played a central role in orchestrating this effort.

\section{Conclusions}
The U.S. CMS internship program is a novel experience where 20 interns get trained into several aspects of CMS research program by several members of U.S. CMS collaboration making this a highly unique and diversity-driven program. The detailed feedback about parts of the program  from 2022 has been very valuable to calibrate 2023 edition of the internship. The program builds a strong bond between collaboration and interns and among interns as well, creating a sense of community. The in-person program of 2023 is certainly a far richer experience than the on-line version in 2022. From the 2023 pool of interns, six students were selected for a semester long internship, extending research beyond summer projects. At least one 2022 intern is now in a PhD program in CMS while others are applying, using recommendations from mentors. The prolonged relationship with collaborators makes much-needed reference letters impactful and increase chances of admission in graduate programs.  Subsequently, several interns have presented research work at conferences and it has boosted confidence and presentation skills. The program is well-structured and aligned to address lack of diversity in STEM and HEP and is one of its kind. U.S. CMS will continue to fulfill its diversity plan using this program as a powerful enabler.

 Financial support is gratefully acknowledged from U.S. CMS SPRINT award at Tougaloo College (Santanu Banerjee), Brown University (Ulrich Heintz), University of Puerto Rico Mayaguez (Sudhir Malik), University of Wisconsin- Madison (Tulika Bose), and U.S. CMS Operations program at Fermilab and the University of Nebraska-Lincoln led, respectively, by Lothar Bauerdick and Ken Bloom. The grant numbers are given as follows:

RENEW-HEP: ``U.S. CMS SPRINT - A Scholar Program for Research INTernship" awarded at:
Award Number - DESC0023681 (Tougaloo University)
Award Number - DESC0023651 (Brown University)
Award Number - DESC0023643 (University of Wisconsin Madison)
Award Number - DESC0023680 (University of Puerto Rico Mayaguez)
and 
U.S. CMS Operations grant NSF-2121686 for “U.S. CMS
Operations at the Large Hadron Collider”

\section*{2024 PURSUE team} The U.S. CMS PURSUE internship program is an effort of a large team of dedicated collaborators. We list below their names and contributions. 

\textbf{Software Tutorials Mentors}: Bruno Coimbra (Fermilab), Marco Mambelli (Fermilab), Shivani Lomte (University of Wisconsin-Madison), Guillermo Fidalgo (University of Puerto Rico Mayaguez), Nick Pervan (Brown University).
\\
\textbf{Fermilab LPC support}: Bo Jayatilaka (Fermilab), Kevin Black (University of Wisconsin-Madison), Marguerite Tonjes (University of Illinois - Chicago), Gabriele Benelli (Brown University), David Yu (University of Nebraska- Lincoln).
\\
\textbf{Cohort buidling, Mentoring, Professional Development}: Julie Hogan (Bethel University), Matthew Bellis (Siena College), Chris Palmer
(University of  Maryland), Cristina Suarez (Fermilab).
\\
\textbf{Project Mentors}: Jennet Dickinson (Fermilab), Jim Hirschauer (Fermilab), Cristina Suarez (Fermilab), Patrick Gartung (Fermilab), Michael Krohn (Fermilab), Nick Smith (Fermilab), Doug Berry (Fermilab), Hans Wenzel (Fermilab), Daniel Diaz (University of California San Diego), Javier Duarte (University of California San Diego), Anthony Aportela (University of California San Diego), Raghav Kansal (University of California San Diego), Jingyu Zhang (Florida State University), Darin Acosta (Rice University), Efe Yigitbasi (Rice University), John Rotter (Rice University), Kiley Kennedy (Princeton University), Andre Frankenthal (Princeton University), Isobel Ojalvo (Princeton University), Sergei Gleyzer (University of Alabama Tuscaloosa), Marcus Hohlman (Florida Institute of Technology), Geamel Alyami (Florida Institute of Technology), Erick Yanes (Florida Institute of Technology), Ana Maria Slivar (University of Alabama Tuscaloosa), Chris Palmer (University of Maryland), Agni Bethani (University of Maryland), Braden Kronheim (University of Maryland), Ulrich Heintz (Brown University), Nick Pervan (Brown University), Abdollah Mohammadi (University of Wisconsin - Madison), Charis Kleio Koraka (University of Wisconsin - Madison), Elise M Chavez (University of Wisconsin - Madison), Tulika Bose (University of Wisconsin - Madison), Kevin Black (University of Wisconsin - Madison), Shivani Lomte (University of Wisconsin - Madison), Abdollah Mohammadi (University of Wisconsin - Madison), Ken Bloom (University of Nebraska - Lincoln), Andrew Wightman (University of Nebraska - Lincoln), Oksana Shadura (University of Nebraska - Lincoln), David Yu (University of Nebraska - Lincoln).
\\
\textbf{Speakers for talks on Detectors, Physics, Computing, Meet-a-Scientist, HEP motivation and organising lab tours, talks by intern alumni}: Tulika Bose (University of Wisconsin - Madison), Julie Hogan (Bethel University), Matt Bellis (Siena College), Steve Nahn (Fermilab), Andrew Melo (Vanderbilt University), Ken Bloom (University of Nebraska - Lincoln), Kevin Black (University of Wisconsin - Madison), Bonnie Flemming (Fermilab), Titas Roy (University of Illinois - Chicago), David Yu (University of Nebraska- Lincoln) Xinyue-Wu (PURSUE Alumni - University of Rochester), Nadja Strobbe (University of Minnesota), Boaz Klima (Fermilab), Daniel Noonan (Fermilab), Don Lincoln (Fermilab), Oliver Gutsche (Fermilab), Sergo Jindariani (Fermilab), Scarlet Norberg (Fermilab), David Velasco (PURSUE alumni - DePaul University), Todd Adams (Florida State University), Douglas Berry (Fermilab), Christian Herwig (Fermilab), Chris Quigg (Fermilab), Robin Erbacher (University of California-Davis), Deborah Pinna (University of Wisconsin - Madison), Sergei Gleyzer (University of Alabama Tuscaloosa), Bogdan Dobrescu (Fermilab), Chris Palmer (University of Maryland), Richa Sharma (University of Puerto Rico Mayaguez), Sudhir Malik (University of Puerto Rico Mayaguez).

\textbf{Webinar support}: Guillermo Fidalgo (University of Puerto Rico Mayaguez)

\textbf{Organisation and logistic support}: Santanu Banerjee (Tougaloo University), Maggie Slusarczyk (Tougaloo University), Tulika Bose (Universty of Wisconsin - Madison), Ulirich Heintz (Brown University), Sudhir Malik (University of Puerto Rico Mayaguez).

\section*{References}

\typeout{}
\bibliography{Main}


\end{document}